\documentclass[superscriptaddress,preprint, asmmath aps, prl, floatfix]{revtex4-1}

\usepackage{amsmath}
\usepackage{graphicx}
\usepackage{dcolumn}
\usepackage{bm}
\usepackage[usenames,dvipsnames]{xcolor}
\usepackage[%
    colorlinks=true,
    pdfborder={0 0 0},
    linkcolor=cyan
]{hyperref}

\usepackage{indentfirst}

\definecolor{red}{rgb}{1,0.,0}

\begin{document}

\title{Simulating the Shastry-Sutherland Ising Model \\ using Quantum Annealing}

\author{Paul Kairys}
\affiliation{Bredesen Center for Interdisciplinary Research and Graduate Education, University of Tennessee, Knoxville, Tennessee, USA}
\affiliation{Quantum Computing Institute, Oak Ridge National Laboratory, Oak Ridge, Tennessee, USA}

\author{Andrew D.~King}
\author{Isil Ozfidan}
\author{Kelly Boothby}
\author{Jack Raymond}
\affiliation{D-Wave Systems, Inc., Burnaby, British Columbia, Canada}

\author{Arnab Banerjee}
\email{arnabb@purdue.edu}
\affiliation{Neutron Sciences Division, Oak Ridge National Laboratory, Oak Ridge, Tennessee, USA}
\affiliation{Department of Physics and Astronomy, Purdue University, West Lafayette, Indiana, USA}

\author{Travis S.~Humble}
\email{humblets@ornl.gov}
\affiliation{Bredesen Center for Interdisciplinary Research and Graduate Education, University of Tennessee, Knoxville, Tennessee, USA}
\affiliation{Quantum Computing Institute, Oak Ridge National Laboratory, Oak Ridge, Tennessee, USA}

\date{\today}


\maketitle

\par
\textbf{
Frustration represents an essential feature in the behavior of magnetic materials when constraints on the microscopic Hamiltonian cannot be satisfied simultaneously. This gives rise to exotic phases of matter including spin liquids \cite{broholm_quantum_2020}, spin ices \cite{bramwell_spin_2001}, and stripe phases \cite{wang2016strong}. Here we demonstrate an approach to understanding the microscopic effects of frustration by computing the phases of a 468-spin Shastry-Sutherland Ising Hamiltonian using a quantum annealer. Our approach uses mean-field boundary conditions to mitigate effects of finite size and defects alongside an iterative quantum annealing protocol to simulate   statistical physics. We recover all phases of the Shastry-Sutherland Ising model -- including the well-known fractional magnetization plateau -- and the static structure factor characterizing the critical behavior at these transitions. These results establish quantum annealing as an emerging method in understanding the effects of frustration on the emergence of novel phases of matter and pave the way for future comparisons with real experiments.
} 
\par 
Materials simulation using quantum annealing (QA) relies on fundamentally different principles as compared to conventional computing methods. Whereas traditional simulated annealing uses thermal excitations and the ergodic principle to find the energetic ground state \cite{kirkpatrick_optimization_1983}, quantum annealing uses an additional transverse magnetic field as a quantum tuning parameter that drives transitions between quantum states \cite{brooke1999quantum,johnson_quantum_2011}. The dynamics induced by the transverse field allows QA to explore the energy landscape, potentially faster than classical approaches \cite{king2019scaling}. 
\par
We use a quantum annealer realized by a two-dimensional lattice of superconducting qubits that implements a transverse-field Ising model (TFIM) defined by \cite{johnson_quantum_2011}
\begin{align}
    \label{eq:dwave-TFIM}
    H(s) = A(s) \sum_i \sigma^{x}_{(i)} + B(s) \bigg[\sum_i h_i \sigma^{z}_{(i)} + \sum_{\langle i,j \rangle} J_{ij} \sigma^{z}_{(i)} \sigma^{z}_{(j)} \bigg]
\end{align}
where the amplitudes $A(s)$ and $B(s)$ determine the relative strength of the transverse field and Ising terms, and are controlled through the dimensionless parameter $s \in [0,1]$. The amplitudes $A(s)$ and $B(s)$ are smooth functions of $s$ with $A(0) >> B(0)$ and $A(1) << B(1)$. The parameters  $J_{ij}$ and $h_i$ define tunable Ising interactions and a per-qubit tunable longitudinal magnetic field, respectively, which enables the realization of various magnetic systems and lattice geometries within the processor. The available couplers are arranged in a ``Chimera'' graph \cite{brunt_magnetic_nodate}, used by several recent examples to validate materials simulations using QA. This includes the simulation of the three-dimensional (3D) spin-glass transition  \cite{harris_phase_2018} and the demonstration of the Berezinskii-–Kosterlitz-–Thouless (BKT) transition in a two-dimensional TFIM \cite{king_observation_2018}. 
\par
\begin{figure}
    \centering
    \includegraphics[width=0.9\textwidth]{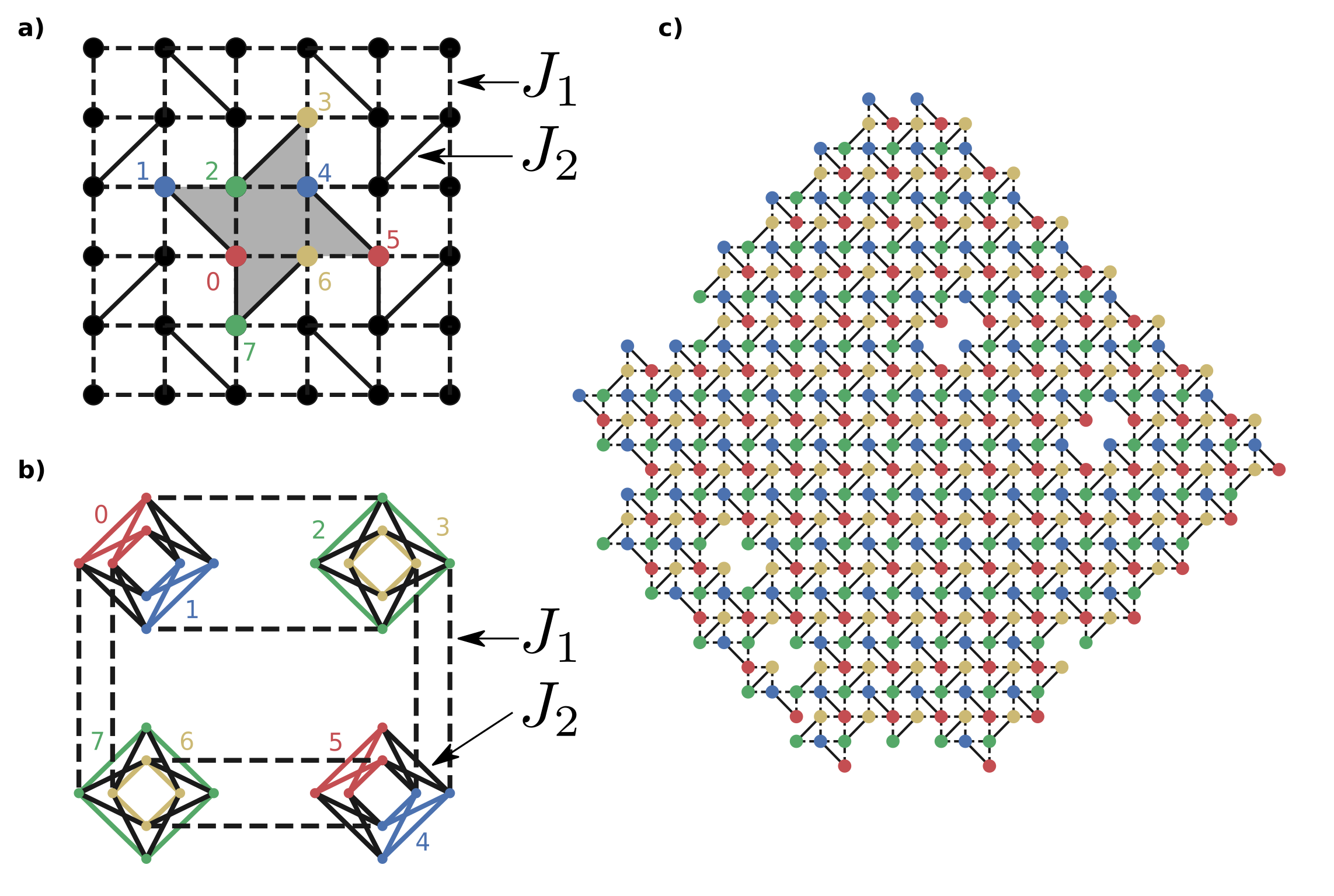}
    \caption{\textbf{Embedding the S-S lattice in the D-Wave quantum annealer.} \textbf{a}, The Shastry-Sutherland lattice showing the nearest neighbor square (dashed black lines) and the next-nearest neighbor dimer (solid black lines) interactions. Highlighted in grey is the logical unit cell with 8 spins labeled. \textbf{b}, The embedding of the logical unit cell from (\textbf{a}) into a Chimera graph (full embedded graph shown in \ref{fig:embedding_figure_appendix}). This ``half-cell'' embedding maps each logical spin to a chain of  4 ferromagnetically connected qubits (solid color lines) and each dimer pair to a single Chimera unit cell. Each logical square bond is mapped to 1 external coupler (dashed black lines) between Chimera unit cells and each dimer bond is mapped to 8 internal couplers (solid black lines) within a Chimera unit cell. This embedding preserves the symmetry of the logical problem and ensures that every physical qubit experiences similar local environments which prevents chain breaking. \textbf{c}, The embedding of a 468-spin lattice in the quantum annealer. Missing lattice sites in the underlying Chimera graph reveal defects that originate from defective qubits or bonds in the annealer. To mitigate both finite size effects and defective sites we apply mean-field boundary conditions. These boundary conditions are applied on both the external boundaries and the internal boundaries (qubits surrounding defects) simultaneously.}
    \label{fig:embedding}
\end{figure}

 
We investigate geometric frustration within an Ising model represented by the Hamiltonian
\begin{align}
\label{eq:ising}
    H = J_1 \sum_{\langle i,j \rangle} \sigma^{z}_{(i)} \sigma^{z}_{(j)} + J_2 \sum_{\langle\langle i,j \rangle\rangle} \sigma^{z}_{(i)} \sigma^{z}_{(j)} + h_z \sum_i \sigma^z_{(i)},
\end{align}
defined over the Shastry-Sutherland lattice \cite{shastry1981exact}, as shown in Fig.~\ref{fig:embedding}(a). In the Shastry-Sutherland Ising model, geometric frustration originates from the nearest-neighbor $J_1$ and the next-nearest neighbor $J_2$ interactions acting on the square frustrated grid. In particular, this model exhibits multiple phases including a trivial ferromagnetic (FM) phase when $h_z \gg J_1,J_2$, a N\'eel anti-ferromagnetic (AFM) phase when $J_1 \gg h_z, J_2$, and a more interesting AFM spin-dimer phase when $J_2 \gg h_z, J_1$ \cite{chang_magnetization_2009}. However, for $J_1 \approx J_2 > 0$, a highly non-trivial phase arises in which spins order to form a 1/3 magnetization plateau. This plateau arises from a 6-fold degenerate solution in the model above a critical longitudinal field strength \cite{chang_magnetization_2009,dublenych_ground_2012}. 
\par
Magnetization plateaus in Shastry-Sutherland models have generated significant interest historically. Notably, the Shastry-Sutherland Ising model and its variations have been proposed to account for the fractional magnetic phases observed in frustrated magnets such as the rare-earth tetraborides (RB$_4$) \cite{dublenych_ground_2012,siemensmeyer_fractional_2008,huang_multi-step_2012}, although there is some debate on the role of additional terms beyond equation~(\ref{eq:ising}). The exact solubility of the Ising model and its simple square topology makes a compelling testbed to evaluate the efficacy of QA for addressing fundamental materials science and exploring a new pathway for the cross-examination of theory and experimental data.
\par 
We use QA to simulate the macroscopic behaviors of the Shastry-Sutherland Ising model by sampling the low-energy states across all phases of the model. Our approach goes beyond conventional QA approaches by using an iterative simulation protocol known as Quantum Evolution Monte Carlo (QEMC) chaining \cite{king_observation_2018,king2019scaling} to sample the ground-state manifold of the frustrated Ising Hamiltonian. We demonstrate this method across the full range of Hamiltonian parameters to recover the complete phase diagram. In order to simulate accurately the bulk behavior, we also apply a technique for mean-field boundary conditions \cite{muller-krumbhaar_1972,landau_guide_2014} to mitigate finite-size effects as well as the presence of defects in the quantum annealer. We then recover both the phase diagram and the static structure factor for the model Hamiltonian from the simulated ensemble of spin states. These results validate the use of programmable quantum annealing as an avenue to understanding frustration in a magnetic Hamiltonian and the potential to simulate the macroscopic behaviors of bulk materials suitable for future experimental comparison.

\par
The geometry of the Shastry Sutherland Ising model was embedded into 1,872 superconducting qubits in a Chimera graph on a D-Wave 2000Q quantum annealer \cite{bunyk_architectural_2014}. The embedding of the Hamiltonian into the processor uses strongly-coupled ferromagnetic chains to encode each logical spin site \cite{choi_minor-embedding_2008}. The embedding is described in Fig.~\ref{fig:embedding}b, where each logical spin maps to a cyclic chain of 4 physical spins, each logical dimer bond to 8 physical internal couplers, and each logical square bond to one physical external coupler. This so-called "half-cell" embedding exploits the symmetry of the Shastry-Sutherland lattice to ensure dimer couplers map to internal couplers and square couplers map to external couplers. The largest model obtained with the half-cell embedding has 468 logical spins as shown in Fig.~\ref{fig:embedding}c.
\par
\par 
The simulation of the embedded Hamiltonian was susceptible to the defects and finite-size effects shown in Fig.~\ref{fig:embedding}c, which can undermine estimates of bulk properties. We introduced mean-field boundary conditions in the embedded Hamiltonian in order to better approximate the equilibrium statistics of the thermodynamic limit. For an ideal, translationally invariant model, the magnetization at the edge of the embedded lattice reflects the bulk system far from any other interface or boundary. We leveraged independent per-qubit control over the longitudinal magnetic fields on each spin to mimic this expected behavior. The longitudinal field applied to each spin site on the boundary was iteratively adjusted to ensure homogeneous magnetization between interior and boundary qubits (see Methods). Convergence of these boundary fields was obtained using gradient descent methods with a logical constraint that the boundary magnetic fields must match the sign of the uniform bulk magnetic field.
\par
Quantum annealing of the transverse-field Ising model used the Hamiltonian from equation~(\ref{eq:dwave-TFIM}) to control the amplitudes $A(s)$ and $B(s)$. By preparing the quantum annealer in an energetic ground state of $H(0)$, adiabatic evolution from $s = 0$ to $s = 1$  prepares the corresponding ground state of the target Hamiltonian. However, satisfying the adiabatic condition is difficult in practice, and additional complications such as system freeze-out limit the ability of the QA to prepare accurate ground states. 
\par 
By contrast, reverse annealing (RA) prepares an eigenstate of the classical Hamiltonian $H(1)$ as an initial guess to the actual ground state, evolves this state from $s = 1$ to $s = s_{p}$ over the ramp time $t_r$ and then holds the Hamiltonian constant for a pause time $t_{p}$ before annealing back to $s = 1$ over time $t_r$. The final prepared state is then measured to generate a candidate spin state. The QEMC chain method uses a sequence of reverse annealing executions to iteratively refine a candidate solution for the ground state of the target Hamiltonian \cite{king_observation_2018} (see Methods). Each iteration begins with the spin state generated by the previous iteration. This repeated process resembles the Markov chain Monte Carlo method, but fails to be perfectly Markovian due to correlations between iterations. We minimized correlations with the initial state by only accepting samples late in the iterative process as part of the prepared statistical ensemble. 
\par 
\begin{figure}
    \centering
    \includegraphics[width=1.0\textwidth]{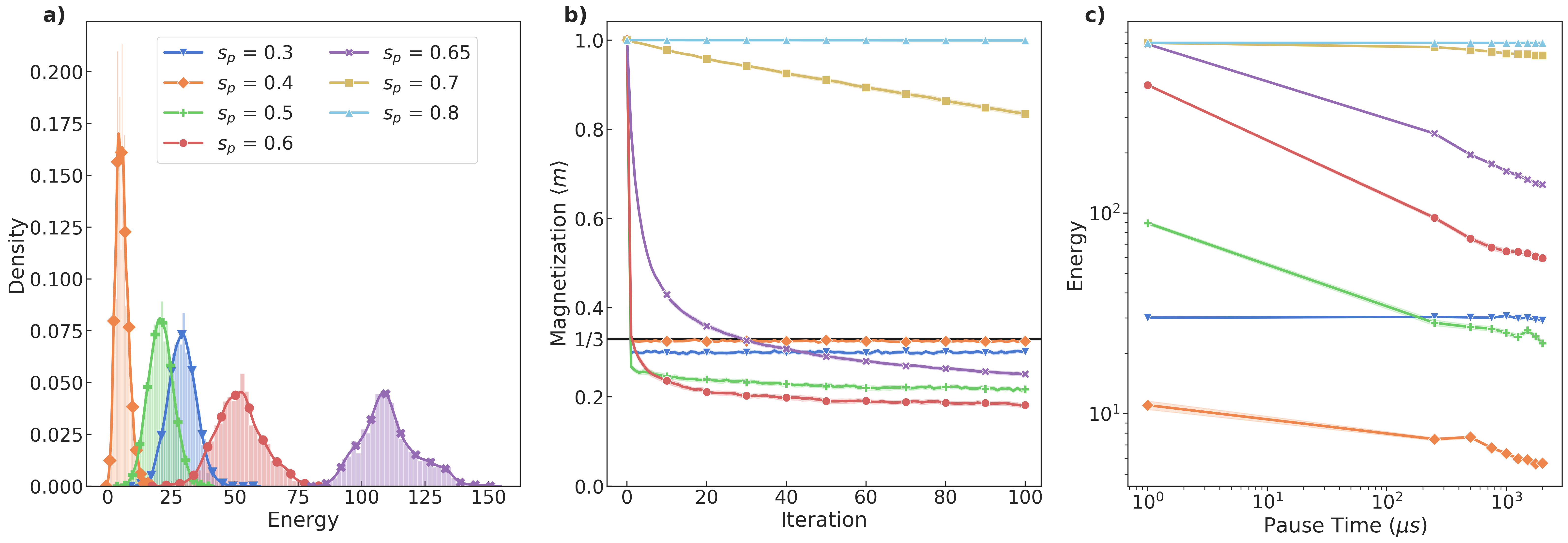}
    \caption{ \textbf{Optimal QEMC annealing parameters.} \textbf{a}, Simulated energy distribution using a chain of QEMCs at $(J_1=1, J_2=1, h=0.5)$ with $t_r = 1~\mu s$ ramps and $t_p = 1998~\mu s$. The energies displayed are re-scaled according to the lowest energy configuration observed in experiment. The legend in (\textbf{a}) is consistent for (\textbf{b}) and (\textbf{c}), but $s_p=0.7,s_p=0.8$ are omitted from (\textbf{a}) for clarity because $\langle H(s_p=0.7) \rangle \geq 600$ and $\langle H(s_p=0.7)\rangle \geq 700$ lie far outside of the domain shown in (\textbf{a}). \textbf{b}, The convergence of the magnetization from a ferromagnetic state at $(J_1=1, J_2=1, h=0.5)$ with $t_r = 1~\mu s$ ramps and $t_p = 1998~\mu s$ to the 1/3 magnetization plateau (shown in black). For $s_p<0.5$ convergence to the bulk magnetization occurs within a few iterations, however these states are not energetically optimal and convergence to the correct microscopic order takes tens of iterations. \textbf{c}, The average returned energy as a function of $s_p$ and $t_p$ with with $t_r = 1~\mu s$. Unless $s_p$ is sufficiently strong such that the timescale of tunneling dynamics is much shorter than the anneal time, increasing pause time can have a dramatic effect in finding low energy solutions.}
    \label{fig:energy_distribution}
\end{figure}
\par 
As shown in Fig.~\ref{fig:energy_distribution}a, we tuned the parameters $s_{p}$ and $t_{p}$ to optimize the prepared ensemble of quantum states. The strength of the transverse field is by the pause position $s_p$, and a value of $s_p \approx 0.4$ returns an ensemble of states with minimal experimental variance and the lowest observed energies over all experiments. For $s>0.4$, we found the solution quality declined rapidly because the transverse field was insufficient to drive tunneling through energy barriers, while for stronger transverse fields ($s<0.4$), we observed the sampling statistics were distorted by the transverse field. As shown in Fig.~\ref{fig:energy_distribution}b, higher transverse field strengths increased the rate of convergence but converged to an ensemble with higher average energy relative to those observed at $s_p=0.4$. As shown in Fig.~\ref{fig:energy_distribution}c, the prepared ensembles were found to be sensitive to the pause duration $t_{p}$ in the range of $1~\mu s$ to $1998~\mu s$ for $s_{p} = 0.4$. By optimizing the annealing parameter $s_p$ and pause time $t_p$, the target magnetization of 1/3 is reached in a single QEMC step. Although this bulk statistic quickly converges, it typically took 40-50 iterations in these simulations to converge to the correct microscopic order.
\par
\begin{figure}
    \centering
    \includegraphics[width=1.0\textwidth]{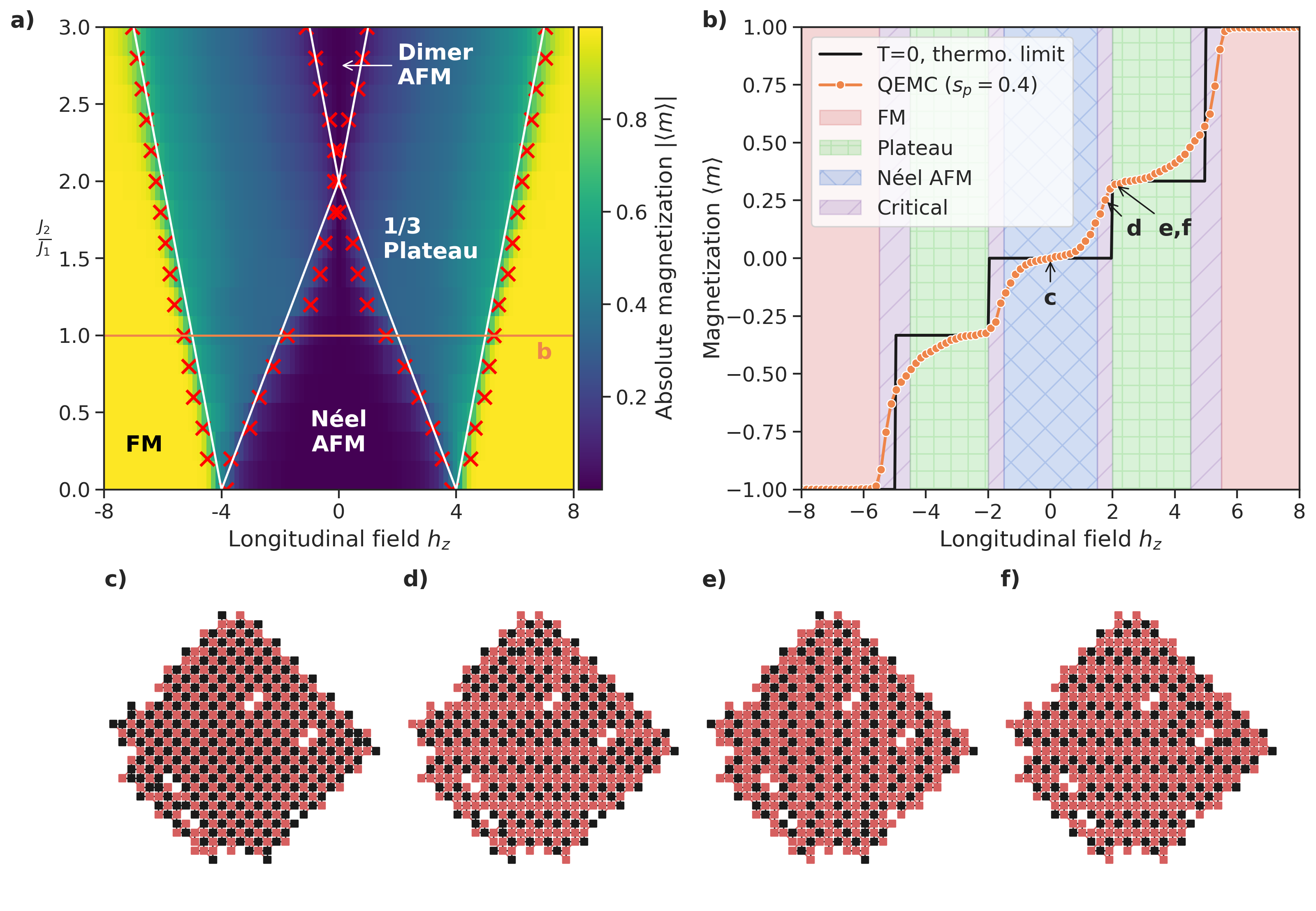}
    \caption{\textbf{Phase diagram and spin structures.} \textbf{a}, Phase diagram obtained by QEMC chains with $t_r $=$ 1~\mu s$, $t_p $=$ 1998~\mu s$ and $s$=$0.4$. The simulated phase boundaries are shown in red by calculating the point at which the derivative of the magnetization is largest which helps to distinguish the 1/3rd plateau, the dimer, FM and the N\'eel states. The white lines outline the exact phase diagram as determined analytically by Dublenych \cite{dublenych_ground_2012}. A slice of the phase diagram at $J_2/J_1 = 1$ shown in (\textbf{b}). \textbf{b}, Slice of phase diagram at ($J_1=1,J_2=1$) obtained by QEMC chains with $t_r $=$ 1~\mu s$, $t_p $=$ 1998~\mu s$ and $s$=$0.4$. The observed phase transition is broadened compared to exact results due to a number of compounding effects, most notably are disorder in $J_1,J_2$, presence of defects and finite size effects and non-negligible persistent transverse field. Four points are labeled c, d, e, and f which correspond to the structures shown in (\textbf{c}-\textbf{f}). \textbf{c}, \textbf{d}, The real-space spin motifs determined by QEMC chains in the N\'eel AFM and critical AFM to plateau phase calculated at the points labeled in (\textbf{b}). In (\textbf{d}) we observe a coexistence between the N\'eel and plateau orderings separated by a domain wall. \textbf{e},\textbf{f}, Two degenerate solutions within the plateau phase. The the black and red squares represent spin up and down, respectively.}
    \label{fig:phase_diagram}
\end{figure}
\par 
We simulated an ensemble of low-energy states for the Shastry-Sutherland Ising model over the range of parameters $h_z \in [-8, 8]$ and $J_2/J_1 \in [0, 3]$, from which we computed the phase diagram shown in Fig.~\ref{fig:phase_diagram}a. 
As indicated by the overlapping theoretical predictions, we obtained excellent agreement with the expected location of the phases and phase transitions across the entire parameter range, reflecting the symmetry of the underlying Hamiltonian. 
Simulation identified all four phases including the $1/3$-magnetization phase, for which a cross-sectional plot is shown in Fig.~\ref{fig:phase_diagram}b. 
The phase transition from the N\'eel anti-ferromagnetic phase to the plateau phase was found at $J_1=1, J_2=1, h_z \approx 1.8$, and the transition from the plateau phase to the ferromagnetic phase was found at $J_1=1, J_2=1, h_z \approx 5$.
\par
In addition to the correct order parameter, we observe the correct microscopic ordering expected in the different phases. In Fig.~\ref{fig:phase_diagram}c we obtain the typical N\'eel AFM ordering with zero average magnetization. In Fig.~\ref{fig:phase_diagram}e,f we display two of the six observed degenerate spin structures found within the plateau phase, each with $\langle m \rangle =1/3$. Shown in Fig.~\ref{fig:phase_diagram}d, at the critical point between N\'eel and plateau phases, we observe intermediate structures with a domain wall between the two coexisting magnetic phases. While these real-space structures are intrinsic to the annealing device, experimental techniques like diffuse neutron scattering of materials do not provide this microscopic information. Moreover, methods from the analysis of diffuse neutron scattering spectra can be utilized here to characterize the spatial correlations in the ensemble of magnetic orders.
\par
\begin{figure}
    \centering
    \includegraphics[scale=0.5]{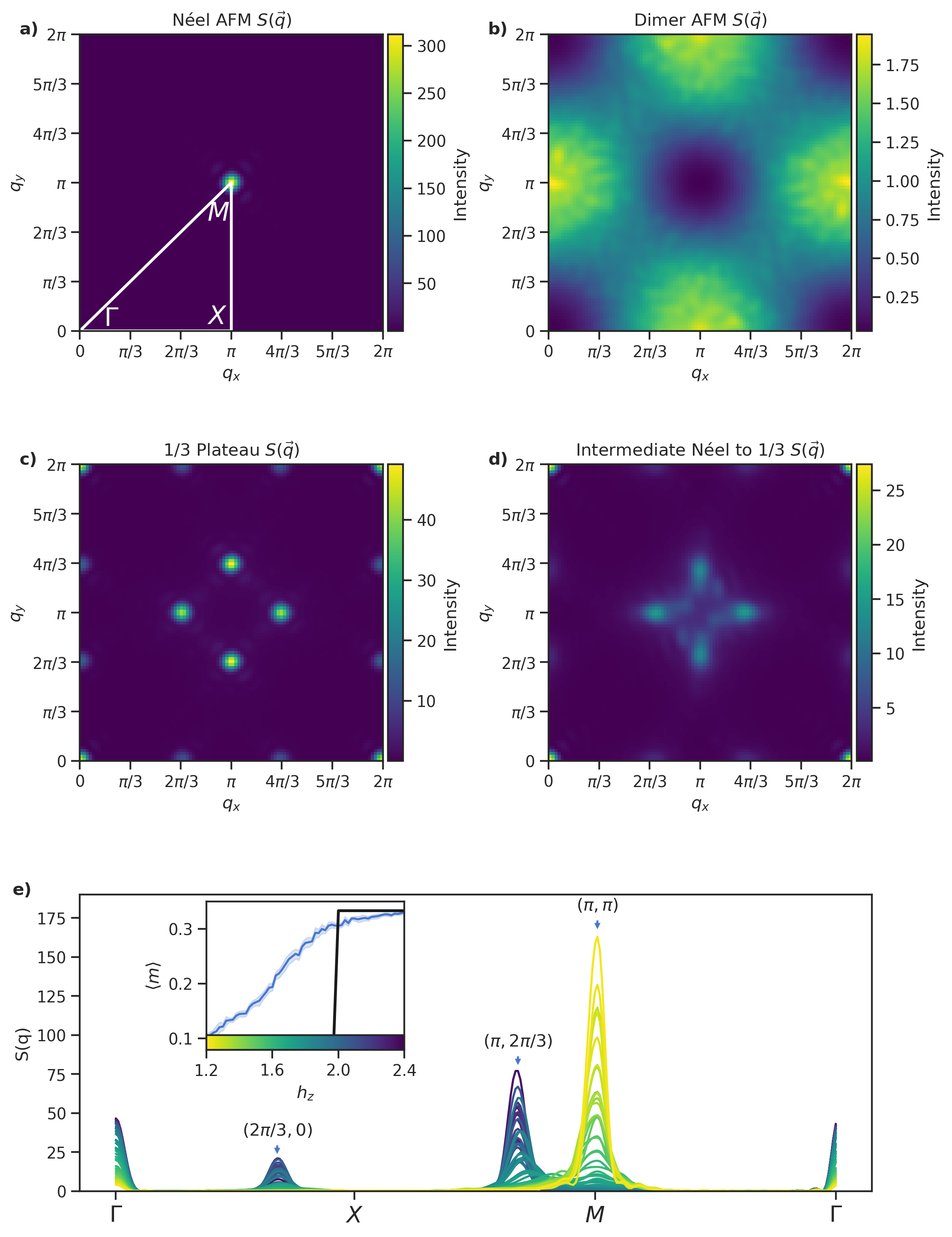}
    \caption{\textbf{Static structure factors in equilibrium and at criticality.} \textbf{a}, Structure factor of the N\'eel AFM phase at $J_1= J_2 = 1, h_z = 0$. The high symmetry points of this model are labeled in white along with a cut-path used in (\textbf{e}), these points are $(q_x=0,q_y=0)$, $(q_x=\pi,q_y=0)$ and $(q_x=\pi,q_y=\pi)$. \textbf{b}, Structure factor of the Dimer AFM phase at $J_1= 1, J_2 = 3, h_z = 0$. \textbf{c}, Structure factor of the 1/3 plateau phase at $J_1= J_2 = 1, h_z = 2.1$. \textbf{d}, Structure factor in the transition between N\'eel AFM and 1/3 plateau phases at $J_1= J_2 = 1, h_z = 1.8$. \textbf{e}, A cut along the symmetry points shown in (\textbf{a}) of the structure factor through the N\'eel to plateau transition. The inset figure shows the color scale corresponding to the longitudinal field across the phase transition at $J_2/J_1=1$.}
    \label{fig:structure_factor_figure_4}
\end{figure}

In order to quantify the spatial correlations, we calculated the static structure factor $S(\vec{q})$ defined in terms of the two-point correlation function as
\begin{equation}
    \label{eq:structure_factor}
    S(\vec{q}) = \sum_{i,j} \langle \sigma^z_{(i)} \sigma^z_{(j)} \rangle e^{i\vec{q} \cdot (\vec{R}_{i}-\vec{R}_{j})}
\end{equation}
where $\vec{R}_{ij}$ is the relative position of two spins and $\vec{q}$ is the wave vector in reciprocal space. The static structure factor provides a Fourier decomposition of the spatial correlations encoded by the spin system and quantifies the ordering of different magnetic phases. The calculated correlations for the N\'eel AFM, dimer AFM, and 1/3 plateau phase are shown in Fig.~\ref{fig:structure_factor_figure_4}a-(c) and agree with theoretical expectations \cite{dublenych_ground_2012}. Because the static structure factor is directly measured by the neutron diffuse scattering spectrum, we anticipate similar QA results for appropriately modified Hamiltonians will enable comparisons with real Shastry-Sutherland magnets in the future. 
\par 
Unlike the standard forward-annealing use of QA, QEMC chains allow us to probe ordering and statistical convergence near critical regimes \cite{king2019scaling} including, for example, the first-order transition from the N\'eel AFM to the 1/3 plateau. However, while theory predicts a sharp discontinuity in the magnetization, our annealing simulations produce a more gradual change as shown in Fig.~\ref{fig:structure_factor_figure_4}e. This smoothing has many origins, including disorder in $J_1,J_2$, defects and finite size effects, as well as non-negligible transverse-field effects persisting near the phase transition. All of these compounding effects lead to the observation of intermediate spin structures which do not exist in either phase. 
\par
While sweeping through the critical regime yielded a smoothed interpolation between the N\'eel and plateau structures, examining the structure factor at a single point in this transition enables us to better visualize the intermediate spatial correlations. Specifically, the structure factor shown in Fig.~\ref{fig:structure_factor_figure_4}d reveals an anisometric broadening of the peaks in the static structure factor that are reminiscent of phenomena seen experimentally in continuous phase transitions \cite{feng_order_2012}. These results underscore the ability of QA methods as computational tools to probe novel environments and suggest new paths to explore the physics in these devices and possible material analogs. 

\par
We have demonstrated how QA may be used to investigate the microscopic origin of frustration in the phases of a model magnetic material. Using QEMC methods with novel boundary conditions, we simulated the low-energy manifold of the Shastry-Sutherland Ising Hamiltonian over a range of model parameters. By accurately recovering the complex phase diagram for this model system as well as the static structure factors for each phase, our results indicate that QA can provide a powerful utility for understanding the emergence of macroscopic behavior from statistical physics.
This work indicates that simulations using a quantum annealer of more complex Hamiltonians and those of real materials is no longer a matter of a conceptual, but rather a technical, challenge. Some of these challenges may be addressed by improved quantum annealing hardware as well as more sophisticated methods of materials simulation \cite{boothby_next-generation_2020}.

\setcounter{figure}{0}
\renewcommand{\thefigure}{S\arabic{figure}}
\section*{Methods}
\begin{figure}
    \centering
    \includegraphics{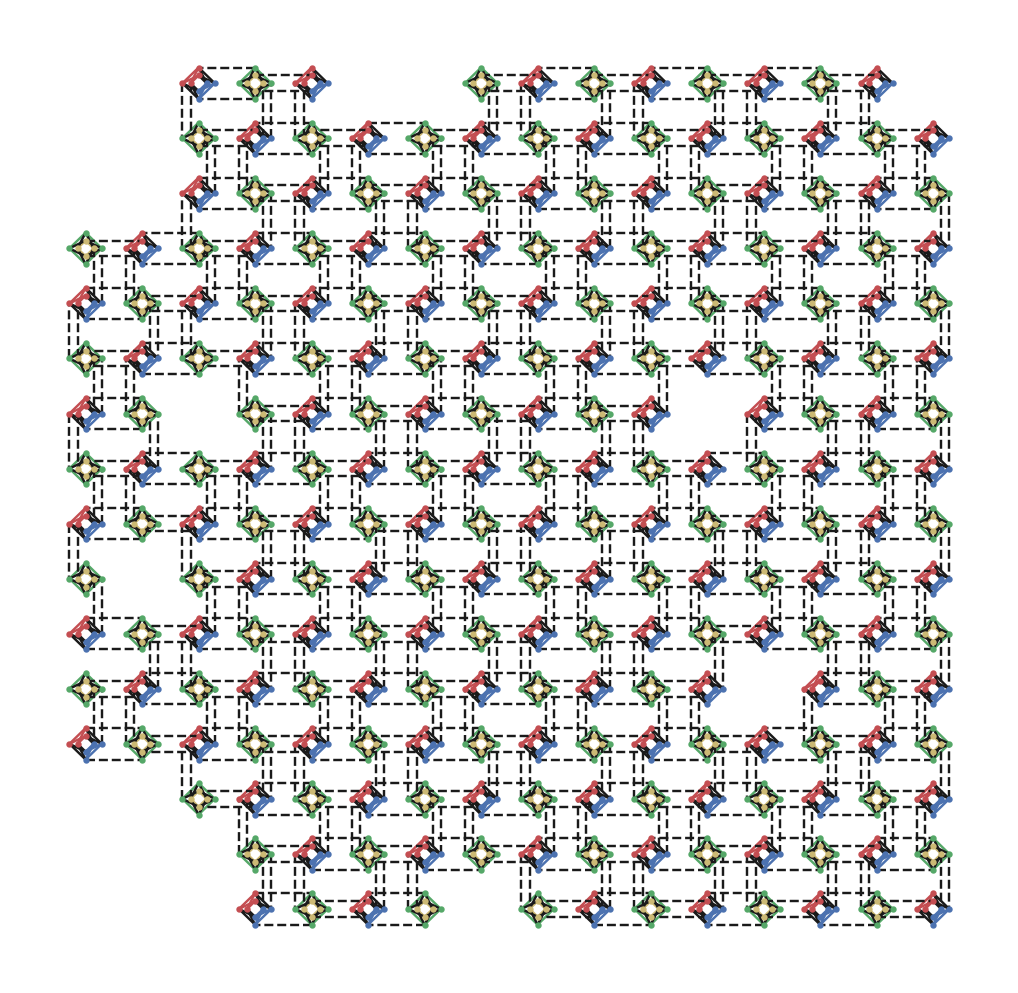}
    \caption{Half-cell embedding into the D-Wave quantum annealer with unit cell as depicted in Fig.~\ref{fig:embedding} (b)}
    \label{fig:embedding_figure_appendix}
\end{figure}

The D-Wave 2000Q quantum annealer is constructed of superconducting flux-qubits coupled together in the Chimera topology \cite{bunyk_architectural_2014}. The quantum annealer has two types of couplers: internal unit cell couplers, which control interactions within a unit cell, and external couplers, which mediate interactions between unit cells. Here we present an embedding, shown in Fig.~\ref{fig:embedding} and Fig.~\ref{fig:embedding_figure_appendix}, which places one logical dimer onto each Chimera unit cell and realizes square bonds between dimers through inter-cell Chimera couplers. This embedding leverages the natural structure of the quantum annealer architecture to reinforce the symmetries of the logical model.

\subsection*{Chi-compensation}

The radio-frequency SQUID flux qubits in the D-Wave 2000Q system do not perfectly implement the TFIM Hamiltonian. Specifically, each qubit mediates an effective coupling between neighboring qubits regardless of the existence of a physical coupler. Additionally there is leakage of an applied $h$ bias from a qubit to its neighbors. The strength of these additional effects are dependent on the normalized background susceptibility $\chi_b = M_{AFM} \chi_q$ where $M_{AFM}$ is the maximum available AFM mutual inductance and $\chi_q$ is the physical qubit susceptibility. These effects lead to a first-order modification to the embedded $h$s and $J$s as shown below.

\begin{align}
    \label{eq:chi_mods_physical}
    J_{i,k}(\chi_b) &= J_{i,k} + \sum_{j} \chi_b J_{i,j} J_{j,k} \\
    h_i(\chi_b) &= h_i + \sum_j \chi_b J_{i,j} h_j
\end{align}

For the low-noise D-Wave 2000Q system used in this work, $\chi_b$ varies as a function of the anneal parameter $s$, for the regime of $0.4<s<0.6$ probed $\chi_b = -0.03 \pm 0.01$. Along with the $J_1, J_2, h$ parameters $J_3$ is the strength of the FM chains that describe a single logical spin, and is fixed in all experiments to $J_3=-1$. For the half-cell embedding presented here it was sufficient to derive expressions of the impact of these interactions on the logical problem, which are given below. We used these expressions to determine the correct input such that the embedded system corresponded to the correct logical problem instance. 

\begin{align}
    \label{eq:chi_mods_logical}
    J_2 (\chi_b) &= J_2 (1+4 \chi_b J_3) \\
    J_1 (\chi_b) &= J_1 (1+4 \chi_b J_3 + 2 \chi_b J_2) \\
    h (\chi_b) &= h (1+\chi_b J_1 +2 \chi_b J_2 + 2 \chi_b J_3) 
\end{align}

\subsection*{Flux-bias offset calibration}

The D-Wave 2000Q system is calibrated as a quantum annealer to perform well under a variety of input types, and the unique problem here permits the re-calibration of the device for improved performance. This calibration should be done to reinforce the expected symmetries in the system, and in the absence of an applied longitudinal field the system is invariant under a spin-flip operation. Using this symmetry we determine flux-bias offsets for each qubit with a gradient descent method such that the average qubit magnetization is zero. The measurement of this magnetization is performed via forward anneals in this work but is generally a function of anneal schedule. 

\subsection*{Mean-field boundary conditions}

Due to the nature of the half-cell embedding, any missing qubit or coupler on the D-Wave 2000Q chip leads to a missing qubit or coupler in the logical SS graph, respectively. Due to the finite size of the chip and the presence of these defects appropriate boundary conditions much be chosen in order to recover the expected solutions in the thermodynamic limit. 

Here we implement a form of mean-field boundary conditions \cite{muller-krumbhaar_1972,landau_guide_2014} which are found via an optimization over the longitudinal magnetic fields of qubits on external and internal boundaries. Specifically, we seek to determine the appropriate magnetic fields such that the magnetization of boundary qubits is equivalent to the magnetization of bulk qubits and is defined as

\begin{equation}
    \begin{aligned}
        \min_{\vec{h}_{bound.}} \quad & \langle m \rangle_{bulk} - \langle m \rangle_{bound.} \\
        \textrm{s.t.} \quad & sign(\vec{h}_{bound.}) = sign(h_{bulk}), \\
    \end{aligned}
\end{equation}
where $\vec{h}_{bound.}$ is the vector of longitudinal magnetic fields on the boundary and $\langle m \rangle_{bulk}$, $\langle m \rangle_{bound.}$ are the magnetization of the bulk and boundary qubits, respectively. The sign constraint is used to restrict the boundary fields to physically realistic local environments. 

The optimization is done via a simple gradient descent method that updates each boundary longitudinal field according to 

\begin{align}
    h^{t+1}_i = h^{t}_i + \delta h (\langle \sigma^z_{(i)} \rangle - \langle m \rangle_{bulk}),
\end{align}

where $\langle \sigma^z_{(i)} \rangle$ is the average magnetization of qubit $i$ over all samples and $0 \leq \delta h \leq 0.05$ is the step size. We typically see convergence of this parameterization up to statistical noise in at most 500 iterations.

\subsection*{Quantum Evolution Monte Carlo Chains}

\begin{figure}
    \centering
    \includegraphics[width=1.0\textwidth]{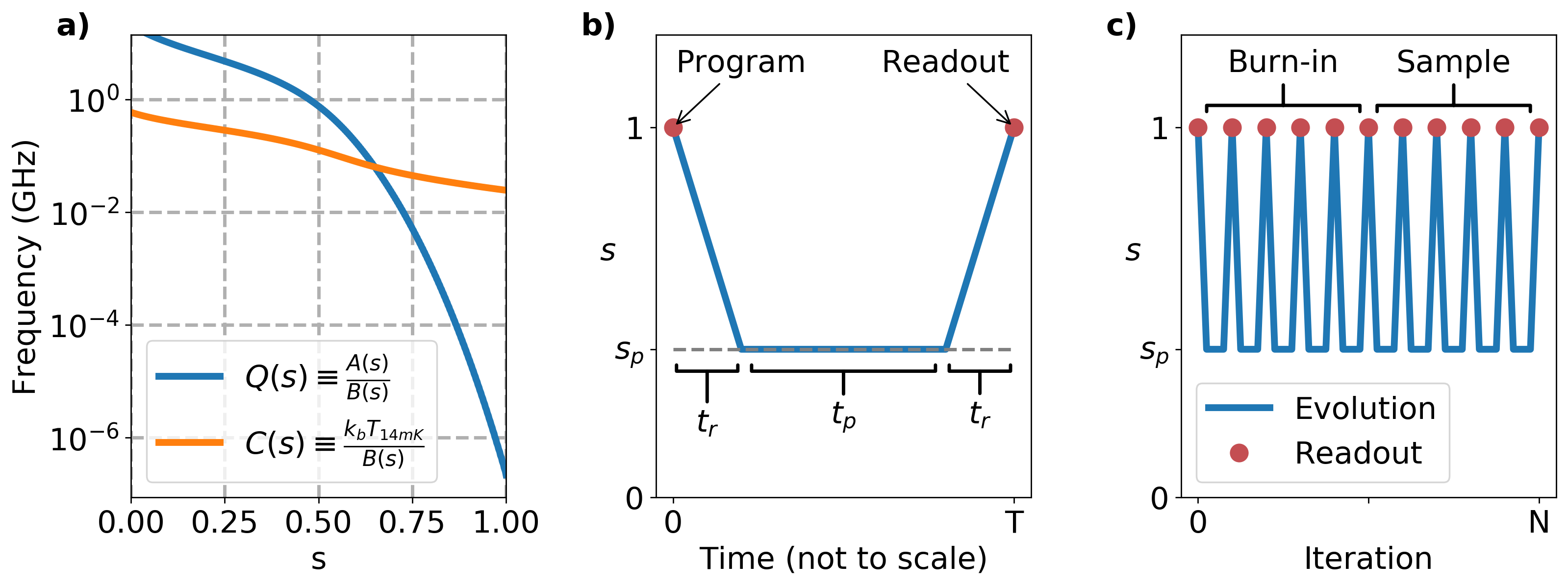}
    \caption{(a) The annealing parameter $s$ controls amplitudes $A(s)$ and $B(s)$ in the device Hamiltonian, equation~\ref{eq:dwave-TFIM}. $Q(s)$ quantifies the relative strengths of these terms as a function of $s$. Also shown is the dependence on $s$ of the ratio to amplitude $B(s)$ and the characteristic thermal energy of the system at the operating temperature of 14 mK. (b) A single Quantum Evolution Monte Carlo (QEMC) anneal schedule that shows the initialization of the device with an eigenstate of the classical Hamiltonian ($A(s)<<B(s)$) at $t=0$, reverse annealing for time $t_r$, pausing for time $t_p$ at $s_p$ (where $Q(s) \geq 1$), and forward annealing for time $t_r$. (c) An iterative protocol where multiple QEMCs are chained together by re-programming the next QEMC with the output from the previous QEMC. The first half of the samples are discarded as burn-in steps while the chain equilibrates and the last half of the samples are collected and used for analysis.}
    \label{fig:qemc_sup}
\end{figure}

In quantum annealing the goal is to evolve a system adiabatically from an eigenstate of the transverse field at $s=0$ to an eigenstate of the Ising model at $s=1$. However for systems composed of chains of ferromagnetically coupled qubits there may be a value of $s$ at which the spins of the chain fail to evolve meaningfully and freeze \cite{amin_searching_2015}. This freeze out point is a function of the average length of the chains and for chains of 4-qubits the freeze out point for this device is approximately $s=0.4$, where the transverse field is non-negligible. If the system freezes near this position in the anneal then the collected measurement statistics will not resemble a classical Ising model but the TFIM. Shown in Fig.~\ref{fig:qemc_sup}(a) is the relative energy scale of the individual terms in the Hamiltonian equation~\ref{eq:dwave-TFIM} as a function of $s$ and their comparison to the characteristic thermal energy of the system.
\par
Here we utilize the method of King et al.~and chain together a sequence of QEMC steps~\cite{king_observation_2018}. In a reverse anneal the system is initiated in a classical state at $s=1$, follows an anneal schedule, and then is read out to obtain a new classical state. The anneal schedule used here is shown in Fig.~\ref{fig:qemc_sup}(b) and is quite similar to that used by King et al.~\cite{king_observation_2018}: rapidly reverse-anneal to a point in the anneal schedule with non-negligible transverse field, pause to allow the system to thermalize and populate low energy intermediate states, and forward anneal to finish the protocol. When repeated, this protocol resembles a Markov chain Monte Carlo and enables the determination of ground states through an iterative refinement of solutions shown in Fig.~\ref{fig:qemc_sup}(c). Just as in a conventional Markov chain simulation, we discard the first half of the iterations as burn-in steps while the system equilibrates and then sample from the last half of the chain. The size of chain length depends greatly on the pause point $s_p$ as can be seen in Fig.~\ref{fig:energy_distribution}(b). 
\par
All experiments performed in this work used the same reverse anneal time, $t_r=1~\mu s$ and used a chain length of 100, so that 50 steps were used for chain equilibration and 50 states are used for statistical sampling. For the system under study the time scale of dynamics is assumed to be much larger than the time scale of the anneal and readout, therefore a rapid forward anneal permits more time to pause and thermalize while not populating high energy states during the forward anneal. Finally, the initial state to begin each QEMC chain in this work was chosen to be the, FM, all spin-up state. We found little dependence of the final state of the chain on the initial state if $s_p$ was small enough such that the transverse field was significantly strong ($s_p \leq 0.6$), as can be seen in Fig.\ref{fig:energy_distribution}(b).

\section*{Acknowledgements}
We thank M.A.~McGuire and J.-Q.~Yan of Oak Ridge National Laboratory for insightful comments on fractional magnetic materials.
This work is supported by the Department of Energy, Office of Science, Early Career Research Program. This research used quantum computing resources of the Oak Ridge Leadership Computing Facility, which is a DOE Office of Science User Facility supported under Contract DE-AC05-00OR22725. Work performed by AB while at SNS is supported by DOE Office of Science User Facilities Division. This manuscript has been authored by UT-Battelle, LLC under Contract No. DE-AC05-00OR22725 with the U.S. Department of Energy. The United States Government retains and the publisher, by accepting the article for publication, acknowledges that the United States Government retains a non-exclusive, paid-up, irrevocable, world-wide license to publish or reproduce the published form of this manuscript, or allow others to do so, for United States Government purposes. The Department of Energy will provide public access to these results of federally sponsored research in accordance with the DOE Public Access Plan. (http://energy.gov/downloads/doe-public-access-plan).

\section*{Author Contributions}
P.K., A.D.K., I.O, A.B., and T.S.H.~conceived the project. A.D.K, I.O., K.B., and J.R. developed the embedding and boundary conditions. P.K., A.D.K., and I.O. implemented and performed simulations and data analysis. P.K, A.D.K, A.B., and T.S.H. analyzed the results and developed conclusions. All authors contributed in writing and editing the manuscript.

\section*{Competing Interests}
A.D.K, I.O., K.B., and J.R. are current employees of D-Wave Systems, Inc. P.K., A.B., and T.S.H. declare no competing interests. 

\section*{Corresponding Authors}
Correspondence to Arnab Banerjee and Travis S.~Humble.

\bibliographystyle{unsrt}
\bibliography{shastry-sutherland}

\end{document}